\documentstyle[11pt,aaspp4]{article}
\begin{document}
\title{NICMOS Observations of Low-Redshift Quasar Host Galaxies\footnote{NOTE: 
this printout contains degraded figures.  Full resolution images can be found in http://www.astro.wellesley.edu/kmcleod/mm.ps}}
\author{K. K. McLeod}
\affil{Whitin Observatory, Wellesley College, Wellesley, MA
02481; kmcleod@wellesley.edu}
\author{B. A. McLeod}
\affil{Harvard-Smithsonian Center for Astrophysics, 60 Garden St., Cambridge, MA 02138; bmcleod@cfa.harvard.edu}

\begin{abstract}
We have obtained Near-Infrared Camera and Multi-Object
Spectrometer images of 16 radio quiet quasars observed as part of a
project to investigate the ``luminosity/host-mass limit.''  The limit
results were presented in McLeod, Rieke, \& Storrie-Lombardi (1999).
In this paper, we present the images themselves, along with 1- and
2-dimensional analyses of the host galaxy properties.  
We find that our model-independent 1D technique is reliable for use
on ground-based data at low redshifts; 
that many radio-quiet quasars live in deVaucouleurs-law hosts,
although some of the techniques used to determine host type are
questionable; 
that complex structure is found in many of the hosts, but that there
are some hosts that are very smooth and symmetric; 
and that the nuclei radiate at $\sim2-20\%$ of the Eddington rate
based on the assumption that all galaxies have central black holes
with a constant mass fraction of 0.6\%.  Despite targeting
hard-to-resolve hosts, we have failed to find any that imply
super-Eddington accretion rates.

\end{abstract}

\keywords{galaxies:photometry---galaxies:active---infrared:galaxies---quasars: general}

\section{Introduction}

\newcommand\cola {\null}
\newcommand\colb {&}
\newcommand\colc {&}
\newcommand\cold {&}
\newcommand\cole {&}
\newcommand\colf {&}
\newcommand\colg {&}
\newcommand\colh {&}
\newcommand\coli {&}
\newcommand\colj {&}
\newcommand\colk {&}
\newcommand\coll {&}
\newcommand\colm {&}
\newcommand\coln {&}
\newcommand\colo {&}
\newcommand\colp {&}
\newcommand\colq {&}
\newcommand\colr {&}
\newcommand\cols {&}
\newcommand\colt {&}
\newcommand\colu {&}
\newcommand\colv {&}
\newcommand\colw {&}
\newcommand\colx {&}
\newcommand\coly {&}
\newcommand\colz {&}
\newcommand\colA {&}
\newcommand\eol{\\}

Host galaxy studies got the opportunity for a real boost in February
1997 when the Near-Infrared Camera and Multi-Object Spectrometer
(NICMOS) was installed on the Hubble Space Telescope (HST). 
NICMOS combines the superb spatial resolution of HST with the
benefits that long wavelengths provide for imaging the redder hosts
against the overwhelming glare of the bluer quasar nuclei.
We have used NICMOS to image 16 radio quiet quasars as part of a
project to investigate the ``luminosity/host-mass limit,'' 
the results of which were presented in McLeod, Rieke, \&
Storrie-Lombardi (1999; hereafter MRS).  In this paper, we present the
images themselves, along with 1- and 2-dimensional analyses of the
host galaxy properties.

The sample, listed in Table \ref{tab-1d}, is composed of all 10
quasars from our ``high-luminosity sample'' (the 26
highest-luminosity PG quasars with $z<0.3$; \cite{mr94b}) that had not
been previously observed with HST.  To this we  
added 6 luminous quasars out to $z=0.4$ for which ground-based
attempts to resolve a host galaxy had failed.  All 16 objects are in
the redshift range $0.13 < z < 0.40$ with an average $z=0.25$.

\begin{deluxetable}{llllllllll}
\scriptsize
\tablecaption{Results of 1D analysis\tablenotemark{a}\label{tab-1d}}
\tablehead{
\colhead{Name} & \colhead{$z$} & \colhead{$m_H$} & \colhead{$M_H$}
& \colhead{$m_H$} & \colhead{$M_H$}  & \colhead{$M_B$} 
& \colhead{$F_{host}/F_{nuc}$}&\colhead{$L/L_{Edd}\tablenotemark{c}$}
\\
& & \colhead{(nuc)} & \colhead{(nuc)}
& \colhead{(host)} & \colhead{(host)\tablenotemark{b}}
& \colhead{(nuc)\tablenotemark{b}}
}
\startdata

PG0026+129 & 0.142 &  13.3 & -25.5 &  15.2 & -23.6 &
-23.7 &  0.17 & 0.16 \\
PG0947+396\tablenotemark{d} & 0.206 &  14.3 & -25.3 &  15.4 &
-24.2 &  -23.1 &  0.37 & 0.05 \\
PG1048+342 & 0.167 &  15.5 & -23.7 &  15.1 & -24.1 &
-23.2 &  1.48 & 0.07 \\
PG1121+422\tablenotemark{d} & 0.224 &  14.6 & -25.3 &
17.0\tablenotemark{e} & -22.8\tablenotemark{e} &  -23.7 &  0.11 & 0.31\tablenotemark{e} \\
PG1151+117 & 0.176 &  14.6 & -24.7 &  15.9 & -23.3 &
-23.6 &  0.29 & 0.19 \\
PG1322+659 & 0.168 &  14.3 & -24.9 &  15.6 & -23.5 &
-23.2 &  0.29 & 0.10 \\
PG1352+183 & 0.158 &  15.0 & -24.0 &  15.1 & -23.9 &
-23.2 &  0.92 & 0.07 \\
PG1354+213 & 0.300 &  15.7 & -24.9 &  16.1 & -24.4 &
-24.6 &  0.73 & 0.16 \\
PG1427+480\tablenotemark{d} & 0.221 &  15.0 & -24.9 &  16.1 &
-23.7 &  -23.3 &  0.35 & 0.11 \\
PG2233+134 & 0.325 &  15.2 & -25.6 &  16.7 & -24.0 &
-24.3 &  0.26 & 0.20 \\
MARK876    & 0.129 &  13.6 & -25.0 &  13.3 & -25.3 &
-23.0 &  1.36 & 0.02 \\
UM357      & 0.334 &  15.1 & -25.8 &  16.7 & -24.0 &
-24.2 &  0.22 & 0.17 \\
Q0530-379  & 0.334 &  15.5 & -25.4 &  16.9 & -23.9 &
-23.4 &  0.28 & 0.09 \\
NAB1612+26 & 0.395 &  15.6 & -25.6 &  17.3 & -23.9 &
-23.4 &  0.22 & 0.10 \\
1628.6+3806\tablenotemark{d} & 0.394 &  15.2 & -26.0 &  16.7 &
-24.4 & -23.7 &  0.26 & 0.08 \\
KUV18217+6419\tablenotemark{d} & 0.297 &  12.3 & -28.2 &  14.7
&-25.7& -26.0 &  0.11 & 0.20 \\
\enddata
\tablenotetext{a}{$H_0=\rm 80~km~s^{-1}~Mpc^{-1},~q_0=0$}
\tablenotetext{b}{Includes k-correction.}
\tablenotetext{c}{Using
$M_B=M_H-2.1-2.5[log_{10}(L/L_{edd})+log_{10}({\Upsilon_{\rm
V}\over{7.2M_\odot/L_\odot}})+log_{10}({f\over{0.006}})-log_{10}({BC\over{12}})]$
with $\Upsilon_{\rm V}, f,~\rm and~BC$ given by their default values
(MRS)}
\tablenotetext{d}{Used PSF0026 instead of quasar's own PSF star}
\tablenotetext{e}{Detection is uncertain}
\end{deluxetable}

We have used a zero point of 21.80 magnitudes in the Vega system for
the F160W filter (\cite{leh00}),
and refer to the resulting magnitudes simply as ``H.''  We note that
this is fainter by 0.31~mag than the value we used in MRS.
For computing rest-frame properties of the hosts, we assume 
$H_0=\rm 80~km~s^{-1}~Mpc^{-1},~q_0=0,~\Lambda_0=0$ throughout, but
our results are  
not strongly sensitive to cosmology.  For example, at $z=0.4$, the
difference in proper distance between $q_0=0$ and 1 is 10\%.  
We also apply galaxy 
H-band k-corrections appropriate for star-forming galaxies, but note
these amount to less than 0.1 magnitude for the redshifts in our
sample.  For the nuclei, we have used colors and k-corrections from
Cristiani \& Vio (1990).  To compute $M_B(nuc)$, we have
assumed that in the B-band all of the light belongs to the nucleus.
This is a reasonable assumption for these high-luminosity quasars
because (i) we know {\it ex post facto} that the galaxies are
generally less luminous than the nuclei in the H-band, and (ii)
the galaxy contribution relative to the nucleus falls dramatically at
shorter wavelengths (see e.g. McLeod \& Rieke 1995, Fig. 1).

\section{NICMOS Observations}\label{sec-obs}

We observed each quasar in a single orbit with the NIC2 camera and the
F160W filter (approximately the H band).  A small amount of time at
the end of each orbit was used to observe a star for characterizing
the point-spread-function (PSF) of the telescope.  These ``PSF stars''
were chosen within 2\arcmin~of the quasar, which is the maximum
allowed slew that does not require overhead for acquisition of new
guide stars.  For 11 quasars, we used our own ground-based H-band
images to locate the infrared-brightest star within the
2\arcmin~limit.  For the remaining five quasars, we chose the
visually brightest star from the Guide Star Catalog.  These
PSF stars do not have the same colors as the quasars, but this has not
caused problems; we show below that the results are insensitive to PSF
used.  

All of the quasars and PSF stars were observed in MULTIACCUM
mode, which records data in a series of increasingly long nondestructive
readouts.  The reduction software determines the average count rate for each
pixel individually by fitting those parts of the integration ramps
that are not saturated.  This allows us to build up an image that is
linear over the entire field, including on the bright quasar core.  
We can therefore avoid some of the difficulties previously encountered
with PSF-subtraction in WFPC2 exposures, where the quasar
core is necessarily saturated out to $\sim0\farcs5$.  The total
on-source times for the quasars ranged from 1792 to 2304 sec, while
the totals for the PSF stars were 32-128 sec.

For most of the quasars, the MULTIACCUM sequences were obtained in a
four-position SPIRAL-DITH pattern with a dither size of 1\farcs0875,
corresponding to 14.5 NIC2 pixels.  This offset was chosen to be large
enough so that the quasar nucleus was on a different set of pixels on
each of the four exposures, but small enough to provide a reliable
half-pixel offset to improve the sampling of the otherwise slightly
undersampled image.  The dithering also allowed recovery of data lost
due to cosmic rays.  The PSF stars were observed in the same 
four-position SPIRAL-DITH, and all of the quasar and PSF dithers were
centered on the same part of the chip (within $<1\arcsec$).  For four
objects with extended hosts 
and/or large companions (PG0026+129, PG0947+396, PG1048+342, MRK876)
we extended the area covered by observing in a 7 or 8 position
SPIRAL-DITH-CHOP pattern, four positions of which coincided with the
positions used for the rest of the objects.

\section{Data Reduction}\label{sec-red}

We reduced all of the quasar and PSF images using {\it nicred\_1.8} , a set
of C programs and IRAF\footnote{IRAF (Image Reduction and Analysis
Facility) is distributed by the National Optical Astronomy
Observatories, which are operated by the Association of Universities
for Research in Astronomy, Inc., under contract with the National
Science Foundation.} 
scripts developed for use with NICMOS images (\cite{bam97}; \cite{leh00}).
This version improves upon earlier versions of the pipeline and {\it
nicred} by lowering the threshhold at which saturation is flagged.
This is especially important for bright sources like our quasar cores,
where significant charge can build up before the first readout.  

The F160W flat and dark images are on-orbit exposures
processed and provided by NICMOS Deputy PI M. Rieke.  For the rest of the
calibration we used the {\it nicred} files except that we modified the
bad pixel mask to account for (i) the coronagraphic hole in the upper
left portion of the images, and (ii) pixels that were obviously bad in
many of our frames.  In addition, cosmic ray hits were flagged by hand
for each image as part of the {\it nicred} reduction process.  

\begin{figure}[tbhf]
\scriptsize
\plotone{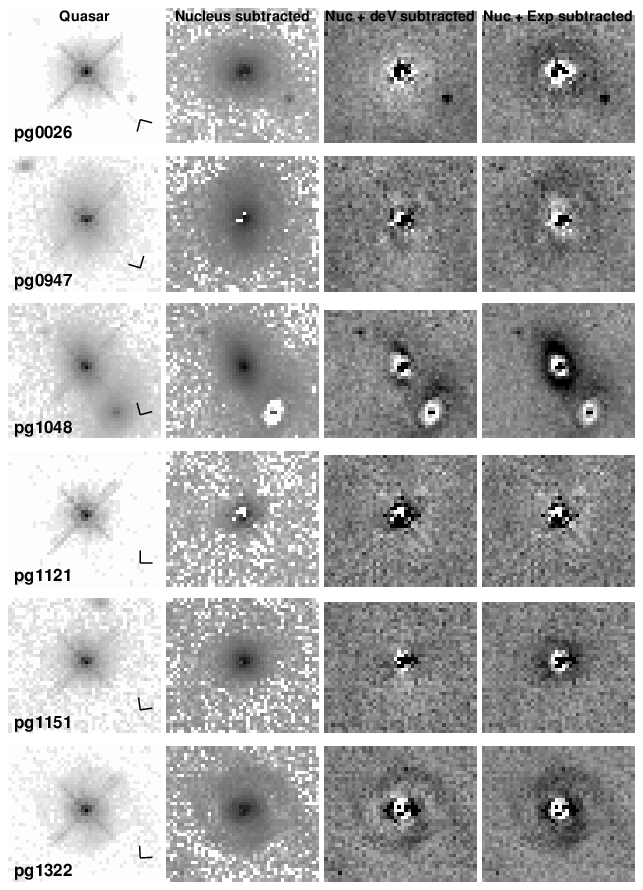}
\figcaption[figure1a.ps]{Central 9\farcs6 of the quasar images on a
logarithmic greyscale stretch.  Angle marker shows North and East
(counterclockwise from N).  From left to right, the columns
are: 
(a) quasar; 
(b) quasar with nucleus removed based on
deVaucouleurs law fit (for KUV18217+6419 we show the exponential law fit
because no unmasked deVaucouleurs model converged);
(c) quasar with nucleus and deVaucouleur model removed;
(d) quasar with nucleus and exponential model removed.  Columns
(b)-(d) were generated from unweighted fits using PSF0026.
\label{fig-qimages}}
\end{figure}

The resulting images from each dither position were magnified by a factor
of 2 and then aligned and combined to produce the final image for each
object.  The central 9\farcs6 of the 20\arcsec~reduced quasar images
are shown in the first column of Fig. \ref{fig-qimages}, and those of the
PSFs are shown in Fig.~\ref{fig-mosaicp}.
The final pixel scale is $\sim0\farcs0375$.  The
PSF ``star'' selected for PG1121+422 turned out to be a compact galaxy
and is not shown.  For the quasar images, the $1\sigma$ scatter per
pixel in blank sky corresponds to a limiting surface brightness of
$H=22.2~\rm mag~arcsec^{-2}$ (or $V\approx25.6~\rm mag~arcsec^{-2}$
for a typical galaxy moved to the average redshift of our sample).

\newpage

\begin{figure}
\plotone{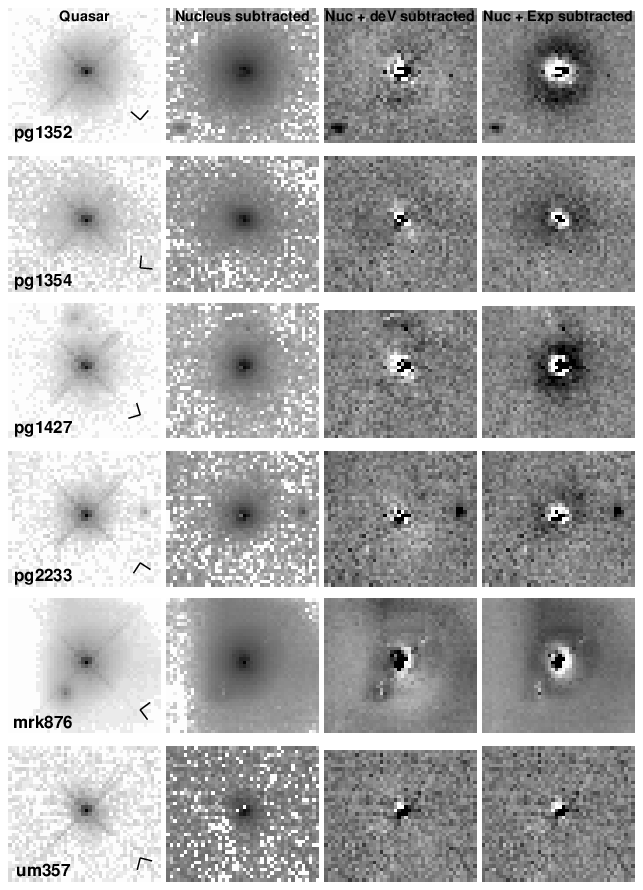}

Fig. \ref{fig-qimages}. --- cont'd
\end{figure}
\begin{figure}
\plotone{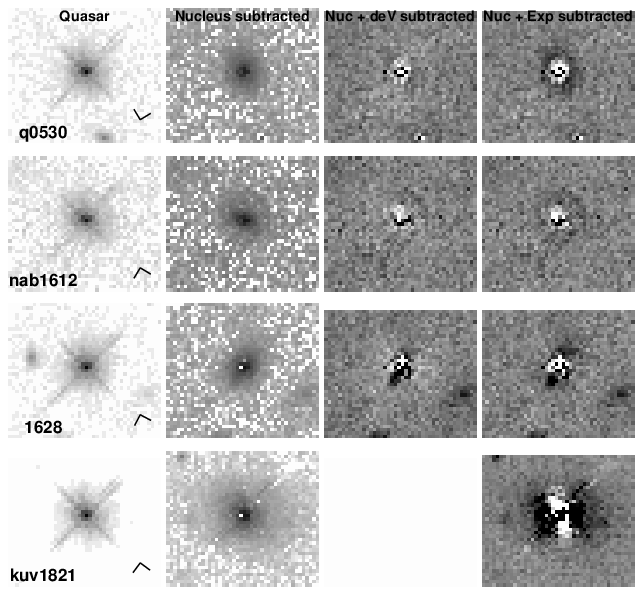}

Fig. \ref{fig-qimages}. --- cont'd
\end{figure}

\begin{figure}[tbhf]
\scriptsize
\plotone{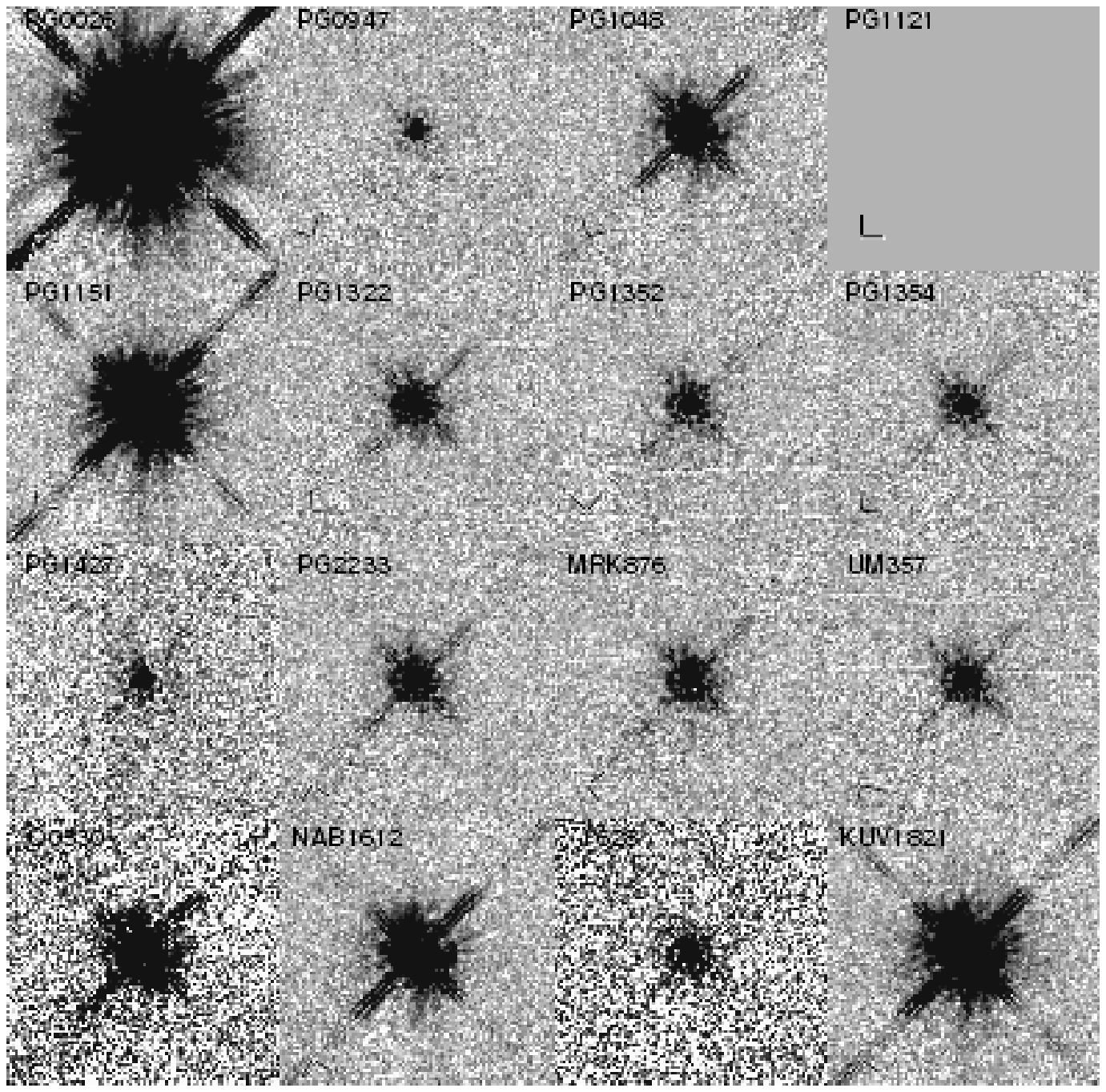}
\figcaption[figure2.ps]{Images of the PSF stars on a logarithmic
greyscale stretch.  Same angular scale and orientations as in
Fig. \ref{fig-qimages}.   
\label{fig-mosaicp}}
\end{figure}

\section{One-Dimensional Analysis}\label{sec-1d}
\subsection{Method}
To estimate the magnitudes of the quasar host galaxies, we first performed a
model-independent 1D removal of the nuclear point source for each
object.  This analysis differs somewhat from the technique used in
MRS, and is better suited to the rereduced the data with the improved
{\it nicred\_1.8} saturation limits.  The results here supersede those
in the earlier paper.
We assume that the H-band light from the quasar is composed of only two
components: the host galaxy, which in the H-band is largely due to
light from red giant stars; and the nucleus, taken to be a point
source with image shape given by the PSF of the telescope.  We do not
assume a particular model for the galaxy, except to assume that its
flux decreases monotonically away from its center.  

\begin{figure}[tbhf]
\epsscale{0.6}
\plotone{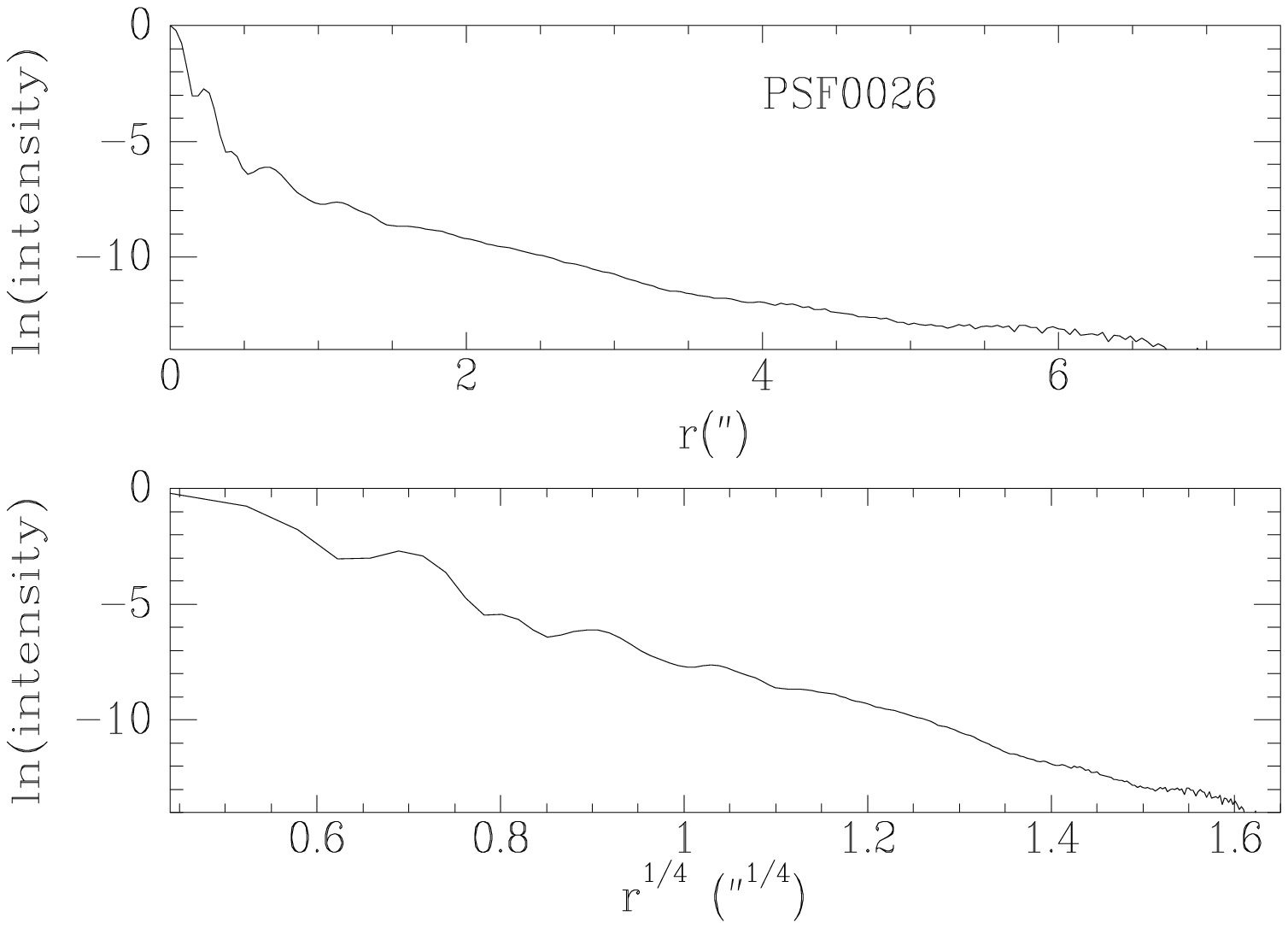}
\figcaption[pg0026_p.eps]{Full 1D radial profile of the brightest PSF
in our sample, PSF0026.  Profiles are shown versus $r$ (top plot) and
$r^{1/4}$ for comparison with quasar profiles in Fig. \ref{fig-profs}.
\label{fig-pg0026_p}}
\end{figure}

\begin{figure}[tbhf]
\epsscale{0.8}
\plotone{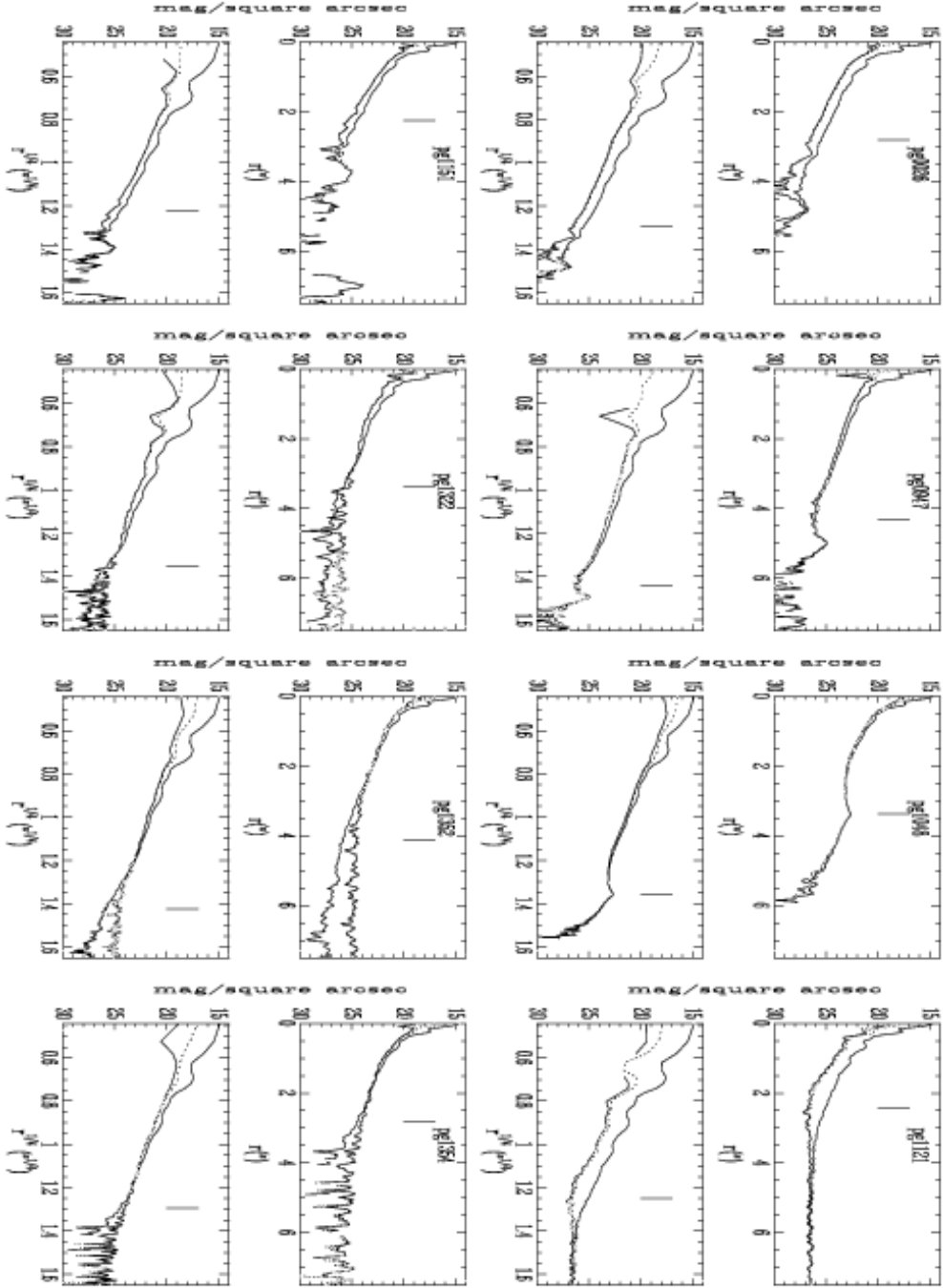}
\figcaption[profsa.eps]{1D radial profiles of the quasars versus $r$
(top plot in each pair) and $r^{1/4}$.  The three curves on each plot
show different PSF subtractions: none (top solid line), 100\% (bottom
solid line), and just monotonic as described in the text (dotted line).  The
vertical line shows the radius to which the profile was integrated
for computation of the host-galaxy magnitude.  Outside of this line,
the profile is compromised by either a companion or noise.  Pure
exponential disks (with no bulge component) would appear as straight
lines versus $r$,, whereas deVaucouleurs galaxies would appear as
straight lines versus $r^{1/4}$.  
\label{fig-profs}}
\end{figure}

We start by extracting from circular annuli 1D radial intensity
profiles of the 16 quasars and the 15 PSF stars.  The profile for the
brightest PSF star, corresponding to PG0026+129 and called hereafter
PSF0026, is shown in Fig. \ref{fig-pg0026_p}.  We then scale the PSF
profiles to have the same central flux as the quasar profile, and
subtract from the quasar the highest PSF fraction that leaves a
monotonic profile inside the first Airy minimum.  Without assuming a
particular functional form for the host galaxy radial profile, this
``just monotonic'' subtraction provides a reasonable estimate of the
host flux (\cite{mr94a}).  Finally, we numerically integrate the
difference profiles, excluding any light from companions.  We can
typically integrate the profiles to a surface brightness of
$H\approx25.7~\rm mag~arcsec^{-2}$ (corresponding to $V\approx29~\rm
mag~arcsec^{-2}$).  The resulting host galaxy magnitudes are given in
Table~\ref{tab-1d}, and the profiles are shown in
Fig.~\ref{fig-profs}.

\begin{figure}
\plotone{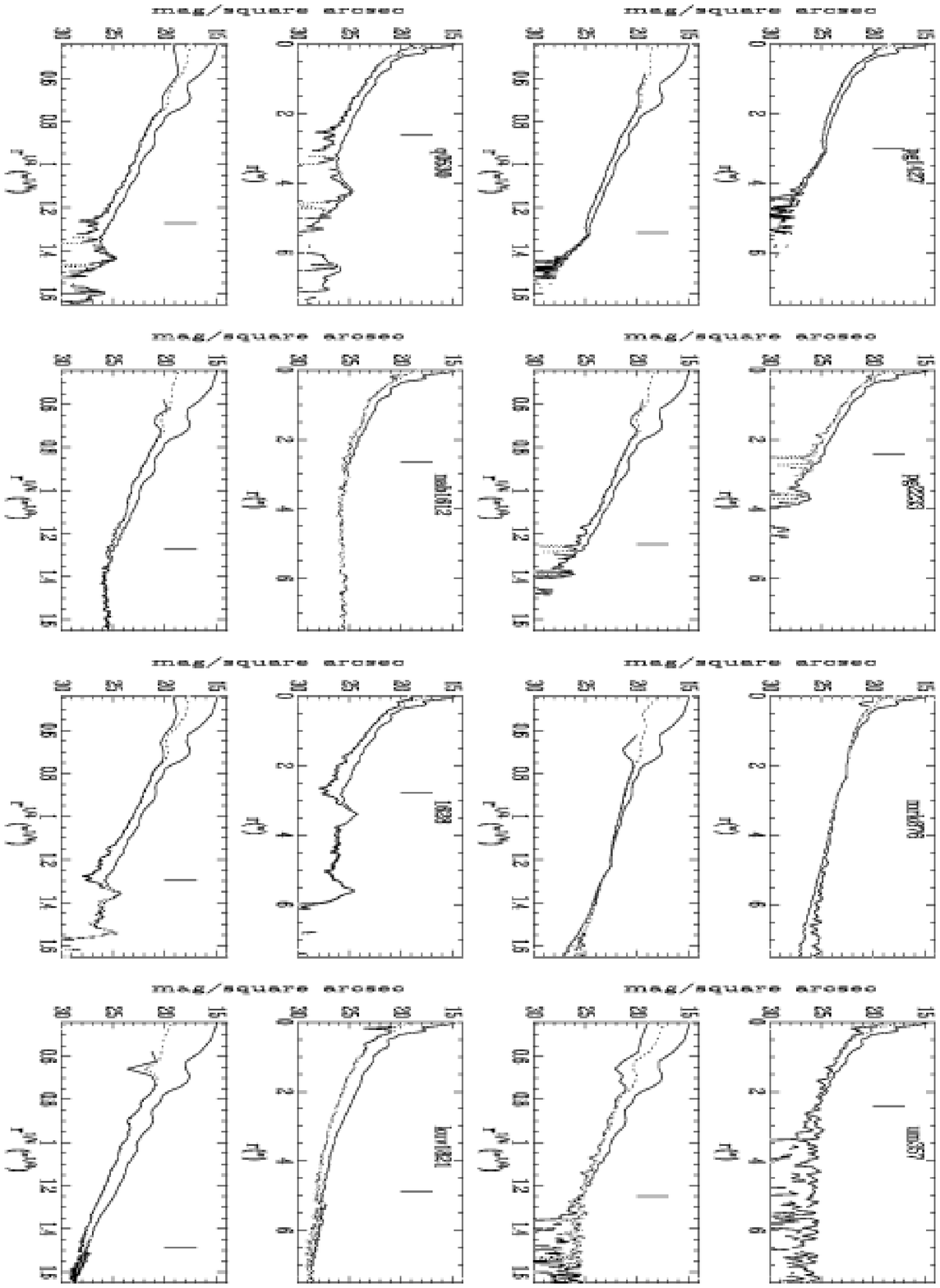}

Fig. \ref{fig-profs}. --- cont'd
\end{figure}

\begin{figure}[tbhf]
\epsscale{0.6}
\plotone{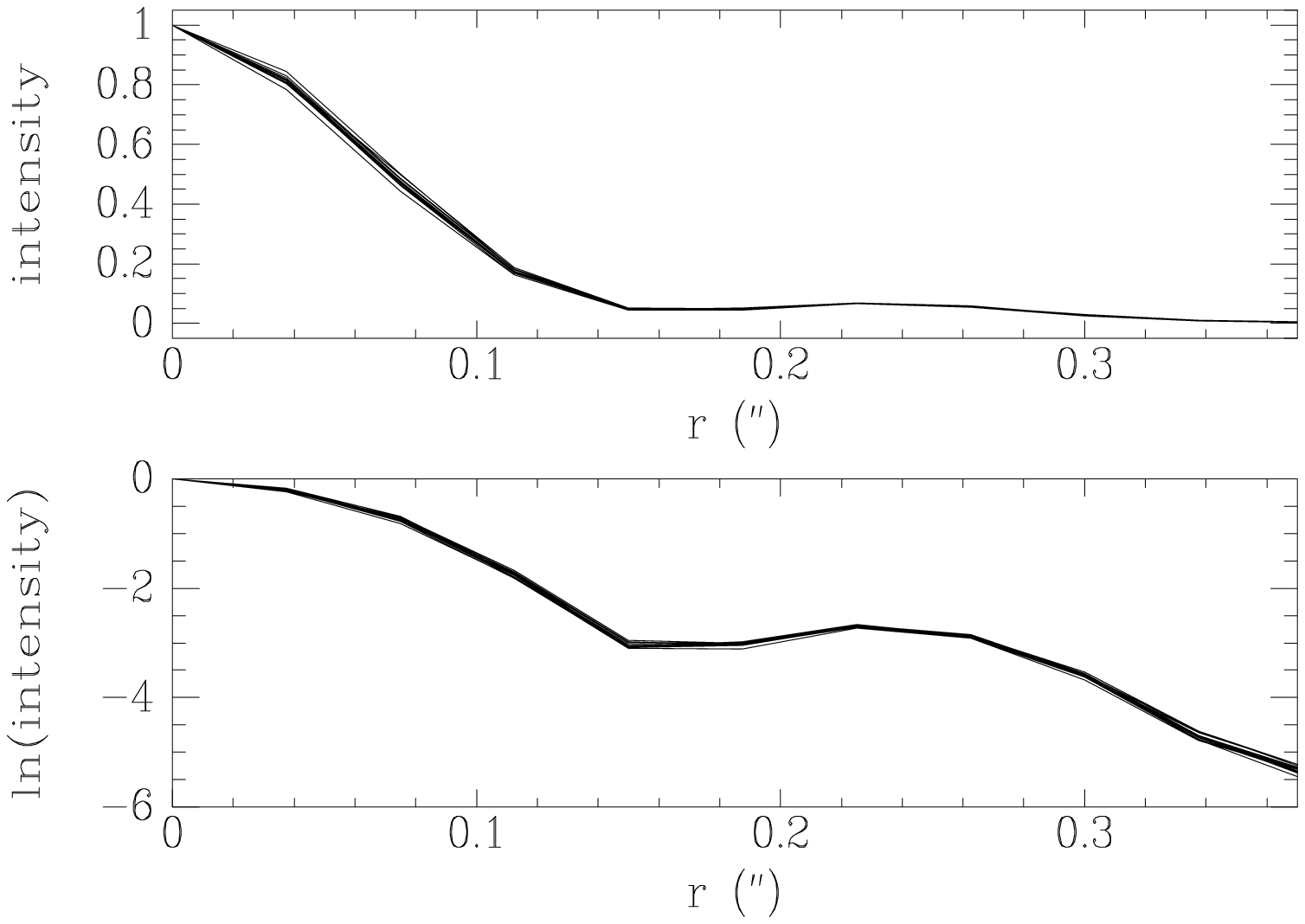}
\figcaption[profs_p.eps]{Central parts of the 1D radial profiles of
all 15 PSF stars, normalized by central intensity.  Profiles are shown
with both linear (top) and natural log scaling to highlight
differences at different radii.  Spreading at large radii is due to
noise in the fainter PSF stars.  
\label{fig-profs_p}}
\end{figure}

In four cases where the quasar's own PSF was faint compared to the
quasar, and in the case of PG1121+422 which has no PSF star, we have used
PSF0026 for the analysis instead.  To test the reliability of using
another quasar's PSF, we repeated our analysis using PSF0026 on all of
the objects.  The resulting host magnitudes were acceptably close, with
$\rm <H(own~psf) - H(PSF0026)>=0.01\pm0.11$.  This is an indication
that the azimuthal average of the NIC2 PSF is relatively stable.  In 
Figure~\ref{fig-profs_p}, we plot the central parts all 15
PSF star profiles together, to show graphically the extent to which
this is so.  

\subsection{Reliability}

The assumption inherent in our 1D method is that the galaxy contributes
little flux to the profile inside the first Airy minimum.  The
procedure we use is effectively equivalent to the one used 
with our ground-based data, where we subtracted just enough PSF to make
the difference profile flatten out at the center.  We can now use
NICMOS data to test the effect of seeing on this procedure for the
ground-based data, by comparing the magnitudes for the  
8 quasars where we have ground-based detections (\cite{mr94b}).  We
find the average value of  $\rm <H(HST) - H(ground)>=0.08\pm0.27$,
which is in good agreement with the 0.30 mag uncertainty quoted for
our ground-based data (\cite{mr94b}).  One of the most discrepant
points is due the presence of a companion that could not be resolved
from the ground.  We conclude that, despite an
order-of-magnitude difference in resolution, the ground-based host
magnitudes derived by this method for low-redshift quasars are fairly
robust when nearby companions can be resolved.  

Our 1D analysis is simplistic but has the advantage of not assuming
{\it a priori} a model for the host galaxy.  The amount by which our
technique underestimates the host flux is dependent on the exact
profile of the galaxy.  To understand the limitations of this method,
we generated a grid of model quasar images covering a range of type
(deVaucouleurs and exponential), effective radius and scale length
(0.5-1\arcsec, corresponding to 1.5-3 kpc for the average redshift of
our sample), and total host intensity relative to the nucleus
(1-30\%, corresponding to galaxies in the range 13-16.7 mag).  We
modeled the quasar as point source + galaxy, and convolved the image
with one our observed PSF stars.  We then ran our 1D analysis
technique on the images using PSF0026 as the ``PSF star.''  We find
that for all the model galaxies with $>1\%$ of the nuclear luminosity,
this technique underestimates the true host flux by $\lesssim0.2$~mag.
For the galaxies at 1\%, errors in estimating the background level for
the profile dominate the procedure, and the resulting host magnitude
errors can be much higher, up to 0.7 mag off in either direction.  For
the real quasars in our sample, the 1D analysis yielded hosts with
total luminosity from 10-150\% of the nuclear luminosity.  Only one
host, that of PG1121+422, is at both a low fraction (11\%) and at the
faint end where background errors are worrisome.  Therefore, we treat
its detection and magnitude as uncertain, but estimate that the rest
of the host magnitudes have an accuracy of $\sim0.2$~mag. 

\subsection{Results}

In Fig.~\ref{fig-profs}, each quasar profile is shown plotted versus
both linear radius $r$ and $r^{1/4}$.  These plots can be used to
judge how well the profiles approximate an exponential disk law
(straight line on the plots versus $r$) or deVaucouleurs law (straight
line on the plots versus $r^{1/4}$).  One can see from this Figure
that the shape of the galaxy's profile outside the central 0\farcs4 is
very little affected by even large errors in PSF subtraction (compare
the ``just monotonic'' and ``100$\%$'' curves).  Eyeball inspection of
these plots shows that some of the profiles clearly follow a
deVaucouleurs law, for example NAB1612+26 and KUV18217+6419.  In some
cases, the choice is dependent upon the PSF fraction subtracted.  We
now turn to a more complete analysis of host morphology using the full
2D information in the images.

\section{Two-Dimensional Analysis}\label{sec-2d}
\subsection{Method}

To characterize the properties of the host galaxies using the full 2D
information in the NICMOS images, we have fit the quasars with
combinations of point sources, deVaucouleurs galaxies, and exponential
disk galaxies, all convolved with a PSF image.  The goodness-of-fit is
determined by a merit that is the weighted sum of the squares of the
residuals over all the pixels.
We modeled each quasar using different combinations of the following
components: (i) the background level (parametrized by a single
number), (ii) the nucleus (intensity $I_n$, position $x_n,y_n$), and
(iii) the host galaxy ($I_g$, $x_g,y_g$, position angle $\theta$, axis
ratio $b/a$, and scale length $r_{eff}$ for a deVaucouleurs law or
$r_0$ for an exponential disk).  When other galaxies were present in
the field, we fit those simultaneously to make sure their
contributions would not bias the host galaxy parameters.

We performed many tests of the fitting procedure's robustness,
finding that we could not always vary all the parameters
simultaneously; in the case of tiny PSF mismatches, the fitting
program would sometimes turn the ``galaxy'' component into a long,
off-center, thin spike to match a high spatial frequency, high
signal-to-noise PSF feature.  We therefore constrained the galaxy and
quasar to have the same position, i.e. we 
assumed the nucleus was in the center of the galaxy.  In tests where
we relaxed this assumption, and where the galaxy was not turned into a
spike, the fitted positions of nucleus and host galaxy agreed within
0\farcs01 ($3\sigma$), corresponding to 30pc at the distance of the
average quasar.

We list in Tables \ref{tab-2d} and \ref{tab-2e} the galaxy parameters
for the deVaucouleurs and exponential models 
from unweighted fits using each quasar's own PSF ($\chi^2$ weighting
is discussed in \S\ref{sec-chisqr}).  The 
host galaxy magnitudes derived  from the deVaucouleurs fits are in
good agreement with the 1D method, with an average difference of $\rm
<H(deV) - H(1D)>=0.07\pm0.27$.  As is commonly seen in host profile
fitting, the exponential fits give hosts systematically fainter than
the 1D fits and deVaucouleurs fits by 0.7~mag, with a $1\sigma$
spread of 0.30~mag. 

\begin{deluxetable}{rrrrrrrrrrrrrrrrrrrrrrr}
\scriptsize

\tablecaption{Results of 2D deVaucouleurs Fits\tablenotemark{a}\label{tab-2d}}

% Table masternic.tab  Mon 15:31:01 07-Feb-2000
\tablehead{
\colb \multicolumn{3}{c}{Own PSF}
\colj \multicolumn{8}{c}{Best PSF} 
\\ \cline{2-4} \cline{5-12}
\colhead{\cola ID} 
\colb \colhead{$m_H$}
\colc \colhead{$r_{eff}$}
\cold \colhead{$b/a$}
\cole \colhead{$m_H$}\colk \colhead{$M_H$}
\coll \colhead{$m_H$}\colm \colhead{$M_H$}
\coln \colhead{$r_{eff}$}\colo \colhead{$r_{eff}$}
\colp \colhead{$b/a$}
\colq \colhead{PSF}
\\
\colb \colhead{(host)} 
\colc \colhead{$(\arcsec)$}
\cold 
\cole \colhead{(nuc)} \colk \colhead{(nuc)}
\coll \colhead{(host)}\colm \colhead{(host)\tablenotemark{b}}
\coln \colhead{$(\arcsec)$} \colo \colhead{$(kpc)\tablenotemark{c}$}
}
\startdata

\cola PG0026+129\colb 14.7\colc   2.3\cold 0.87\cole  13.2\colf -25.6\colg 14.7
\colh -24.1\coli   2.3\colj   4.9\colk 0.87\coll 0026\eol
\cola PG0947+396\colb ...\colc   ...\cold ...\cole  14.3\colf -25.4\colg 14.9
\colh -24.7\coli   2.6\colj  7.4\colk 0.83\coll 1821\eol
\cola PG1048+342\colb 15.5\colc   0.4\cold 0.67\cole  15.2\colf -24.0\colg 15.4
\colh -23.8\coli   0.8\colj  1.8\colk 0.61\coll 0026\eol
\cola PG1121+422\colb ...\colc   ...\cold ...\cole  14.5\colf -25.4\tablenotemark{d}\colg 17.2\tablenotemark{d}
\colh -22.6\tablenotemark{d}\coli   3.9\tablenotemark{d}\colj 11.8\tablenotemark{d}\colk 0.92\tablenotemark{d}\coll 0026\eol
\cola PG1151+117\colb 15.9\colc   0.4\cold 0.66\cole  14.5\colf -24.8\colg 15.7
\colh -23.6\coli   0.6\colj   1.6\colk 0.82\coll 0026\eol
\cola PG1322+659\colb ...\colc  ...\cold ...\cole  14.1\colf -25.0\colg 15.8
\colh -23.4\coli   0.7\colj   1.8\colk 0.95\coll 0026\eol
\cola PG1352+183\colb 15.3\colc   1.2\cold 0.83\cole  14.8\colf -24.3\colg 15.2
\colh -23.8\coli   1.4\colj   3.3\colk 0.79\coll 1821\eol
\cola PG1354+213\colb 16.1\colc   1.0\cold 0.71\cole  15.6\colf -25.0\colg 16.2
\colh -24.3\coli   0.7\colj   2.7\colk 0.75\coll 0026\eol
\cola PG1427+480\colb 16.0\colc   0.5\cold 0.89\cole  14.9\colf -24.9\colg 16.0
\colh -23.8\coli   0.5\colj   1.5\colk 0.89\coll 0026\eol
\cola PG2233+134\colb 16.5\colc   0.6\cold 0.95\cole  15.2\colf -25.6\colg 16.7
\colh -23.9\coli   0.4\colj   1.7\colk 0.63\coll 1821\eol
\cola MARK876\colb 13.4\colc   1.8\cold 0.83\cole  13.3\colf -25.3\colg 13.0
\colh -25.5\coli   4.1\colj  8.0\colk 0.80\coll 0026\eol
\cola UM357\colb 17.1\colc   1.3\cold 0.54\cole  14.9\colf -25.9\colg 17.2
\colh -23.5\coli   1.3\colj   5.0\colk 0.43\coll 0026\eol
\cola Q0530-379\colb 17.3\colc   0.2\cold 0.71\cole  15.3\colf -25.5\colg 17.1
\colh -23.6\coli   0.7\colj   2.7\colk 0.71\coll 0026\eol
\cola NAB1612+26\colb 17.2\colc   0.3\cold 0.79\cole  15.6\colf -25.6\colg 17.2
\colh -23.9\coli   0.3\colj   1.4\colk 0.79\coll 1612\eol
\cola 1628.6+3806\colb 16.7\colc   0.2\cold 0.67\cole  15.1\colf -26.1\colg 16.8
\colh -24.4\coli   0.6\colj   2.6\colk 0.73\coll 0026\eol
\cola KUV18217+6419\colb ...\colc   ...\cold ...\cole  12.3\colf -28.2\colg 14.7
\colh -25.7\coli   0.3\colj  1.1\colk 0.87\coll 1821\tablenotemark{e}\eol

\enddata
\tablenotetext{a}{$H_0=\rm 80~km~s^{-1}~Mpc^{-1},~q_0=0$}
\tablenotetext{b}{Includes k-correction}
\tablenotetext{c}{Based on angular diameter distance, not an isophotal one}
\tablenotetext{d}{Detection is uncertain}
\tablenotetext{e}{From masked fit; no unmasked fits converged}
\end{deluxetable}

\begin{deluxetable}{rrrrrrrrrrrrrrrrrrrrrrr}
\scriptsize

\tablecaption{Results of 2D Exponential Fits\tablenotemark{a}\label{tab-2e}}

% Table masternic.tab  Mon 15:31:01 07-Feb-2000
\tablehead{
\colb \multicolumn{3}{c}{Own PSF}
\colj \multicolumn{8}{c}{Best PSF} 
\\ \cline{2-4} \cline{5-12}
\colhead{\cola ID} 
\colb \colhead{$m_H$}
\colc \colhead{$r_0$}
\cold \colhead{$b/a$}
\cole \colhead{$m_H$}\colk \colhead{$M_H$}
\coll \colhead{$m_H$}\colm \colhead{$M_H$}
\coln \colhead{$r_0$}\colo \colhead{$r_0$}
\colp \colhead{$b/a$}
\colq \colhead{PSF}
\\
\colb \colhead{(host)} 
\colc \colhead{$(\arcsec)$}
\cold 
\cole \colhead{(nuc)} \colk \colhead{(nuc)}
\coll \colhead{(host)}\colm \colhead{(host)\tablenotemark{b}}
\coln \colhead{$(\arcsec)$} \colo \colhead{$\rm (kpc)\tablenotemark{c}$}
}
\startdata
\cola PG0026+129\colb 15.8\colc   0.4\cold 0.81\cole  13.2\colf -25.6\colg 15.8
\colh -23.0\coli   0.4\colj  0.8\colk 0.81\coll 0026\eol
\cola PG0947+396\colb ...\colc   ...\cold ...\cole  14.3\colf -25.4\colg 15.9
\colh -23.7\coli   0.5\colj  1.5\colk 0.77\coll 1821\eol
\cola PG1048+342\colb 16.1\colc   0.2\cold 0.69\cole  15.1\colf -24.0\colg 16.0
\colh -23.1\coli   0.2\colj  0.6\colk 0.64\coll 0026\eol
\cola PG1121+422\colb ...\colc   ...\cold ...\cole  14.5\colf -25.4\colg 18.0\tablenotemark{d}
\colh -21.8\tablenotemark{d} \coli   0.7\tablenotemark{d} \colj
2.0\tablenotemark{d} \colk 0.79\tablenotemark{d} \coll 0026\eol
\cola PG1151+117\colb 16.3\colc   0.2\cold 0.81\cole  14.5\colf -24.8\colg 16.2
\colh -23.0\coli   0.2\colj  0.6\colk 0.90\coll 0026\eol
\cola PG1322+659\colb ...\colc   ...\cold ...\cole  14.1\colf -25.1\colg 16.4
\colh -22.8\coli   0.2\colj  0.6\colk 0.92\coll 0026\eol
\cola PG1352+183\colb 16.0\colc   0.3\cold 0.87\cole  14.7\colf -24.3\colg 15.9
\colh -23.1\coli   0.4\colj  0.9\colk 0.89\coll 0026\eol
\cola PG1354+213\colb 16.8\colc   0.3\cold 0.80\cole  15.5\colf -25.0\colg 16.8
\colh -23.7\coli   0.2\colj  0.9\colk 0.81\coll 0026\eol
\cola PG1427+480\colb 16.5\colc   0.2\cold 0.91\cole  14.9\colf -24.9\colg 16.5
\colh -23.3\coli   0.2\colj  0.5\colk 0.91\coll 0026\eol
\cola PG2233+134\colb 17.1\colc   0.2\cold 0.95\cole  15.1\colf -25.6\colg 17.0
\colh -23.6\coli   0.3\colj  1.0\colk 0.94\coll 0026\eol
\cola MARK876\colb 14.5\colc   0.4\cold 0.88\cole  13.3\colf -25.3\colg 14.3
\colh -24.2\coli   0.5\colj  1.0\colk 0.87\coll 0026\eol
\cola UM357\colb 17.8\colc   0.4\cold 0.72\cole  14.9\colf -25.9\colg 17.8
\colh -22.9\coli   0.4\colj  1.4\colk 0.56\coll 0026\eol
\cola Q0530-379\colb 17.7\colc   0.1\cold 0.74\cole  15.3\colf -25.5\colg 17.7
\colh -23.0\coli   0.2\colj  0.8\colk 0.77\coll 0026\eol
\cola NAB1612+26\colb 17.7\colc   0.2\cold 0.74\cole  15.5\colf -25.7\colg 17.7
\colh -23.4\coli   0.3\colj  1.1\colk 0.60\coll 0026\eol
\cola 1628.6+3806\colb 17.2\colc   0.1\cold 0.73\cole  15.1\colf -26.1\colg 17.2
\colh -23.9\coli   0.2\colj  1.1\colk 0.76\coll 0026\eol
\cola KUV18217+6419\colb ...\colc   ...\cold ...\cole  12.3\colf -28.3\colg 15.3
\colh -25.1\coli   0.3\colj  1.3\colk 0.46\coll 0026\eol
\enddata
\tablenotetext{a}{$H_0=\rm 80~km~s^{-1}~Mpc^{-1},~q_0=0$}
\tablenotetext{b}{Includes k-correction}
\tablenotetext{c}{Based on angular diameter distance, not an isophotal
one}
\tablenotetext{d}{Detection is uncertain}
\end{deluxetable}

In the second column of Fig.~\ref{fig-qimages}, we show the 
quasars after removal of the nucleus based on 2D deVaucouleurs
fits using the high signal-to-noise PSF0026 (at this stretch, the ones
based on 2D exponential fits are visually indistinguishable).
We also show in Fig.~\ref{fig-qimages} the residuals after subtracting both
the nucleus and the model galaxy for both types of host.  The fitting
technique has done a superb 
job of removing most of the complex features of the NICMOS PSF.
It is immediately obvious that few of these hosts are perfectly 
fit by ideal deVaucouleurs and exponential models.  However,
subtracting out the smooth models does reveal interesting structure,
which we discuss in \S\ref{sec-2dresults}.  In general, structure can be
believed outside of the central 0\farcs5 diameter noisy region, and,
in a few galaxies, the regions around the diffraction spike
residuals.  

\subsection{Reliability}

To test the sensitivity of the fits to the properties of the PSF star and to
possible sampling effects in the PSF cores, we repeated the
deVaucouleurs and exponential fits for each quasar with the five PSF
stars having the highest signal-to-noise, running each fit both with
and without a mask that excluded the quasar nucleus inside a 0\farcs37
diameter. Thus, we have fit each quasar $5\times2\times2=20$ different
ways.  We list in Tables \ref{tab-2d} and \ref{tab-2e} the results of
the ``best'' fits, i.e. those using the PSF that gave the lowest merit
for each quasar and galaxy type.  This was often the high
signal-to-noise PSF0026.   We find that there are minor differences in 
the residuals around the first Airy ring, which can
probably be attributed to telescope breathing.  The effect on the
photometry can be quantified by 
comparing the results of the ``own'' and ``best'' fits.  We find that the host
magnitude is robust for both deVaucouleurs and exponential fits, with
$\rm <H(own) - H(best)>_{deV}=-0.04\pm0.18$ and
$\rm <H(own) - H(best)>_{exp}=-0.04\pm0.08$ respectively.  The radii
for exponential fits generally agree within 0\farcs05, corresponding to
0.15~kpc for a typical quasar in our sample.  However, with a scatter
of 0\farcs7, the deVaucouleurs radii are found to be much less certain.
This effect, seen also by McLure, Dunlop, \& Kukula (1999; hereafter
MDK), arises from a degeneracy whereby the fit can steal light from
the nucleus to put into the peaky deVaucouleurs 
galaxy.  For both types of galaxies, the axial ratios agree within
$b/a\sim0.1$.  We find a nearly identical result when we compare the
results from the normal fits to those where we mask out the central
few pixels to avoid having the merit function dominated by noise in
the bright centers. 

As a further test of the 2D method, we explored the level to which our
method might fabricate galaxies when none are detectable.  To do this,
we forced nucleus + galaxy fits to two true point 
sources, namely the bright PSFs corresponding to KUV18217+6419 and
PG1151+117 (a PSF that is not as well matched to the others).  We fitted
them using the 5 PSFs used with the quasars and assuming both
deVaucouleurs and exponential models for the ``galaxies.''  Of
these 20 combinations, the fits diverged to give negative or linear
``galaxies'' 18 times.  The negative fluxes corresponded to a level only
0.4\% that of the point source.  In two cases with PG1151+117's PSF, the method
fabricated an exponential disk with reasonable physical parameters.
These ``galaxies'' have fluxes of only 12\% and 2\% that of the point
source.  Given the flux levels for our quasars, this test
suggests that all the host detections are secure with the possible
exception of PG1121+422 and KUV18217+6419, for which any exponential solutions
must be treated as suspect.  

Finally, the nuclear magnitudes from all of the fits appear extremely
robust.  We measure 
$\rm <H(deV) - H(exp)>=0.025\pm0.021$,
$\rm <H(deV) - H(1D)>=-0.13\pm0.10$, and
$\rm <H(exp) - H(1D)>=-0.15\pm0.09$
The three quasars with large differences ($\sim 0.3~\rm mag$) in nuclear
magnitudes between the 1- and 2D fits are the ones in which the
galaxy light constitutes $\gtrsim0.5$ of the total in the H-band. 

We conclude based on these tests that, for pure exponentials and
deVaucouleurs hosts, the nuclear magnitudes and the
host magnitudes, position angles, and exponential scale lengths are
robust within the uncertainties quoted above, but that the
deVaucouleurs radii are less reliable.

\subsection{$\chi^2$ Weighting}\label{sec-chisqr}

Based on inspection of the residual images and the tests described
above, we believe our unweighted fits give good estimates of the host
parameters.  However, the merit function used does not provide a
well-defined statistical discriminant between deVaucouleurs and
exponential models for the hosts.  Following the approach of MDK, we
have attempted to assess the preference for one or the other type
using a $\chi^2$ analysis.   

We first generated an azimuthally symmetric error map from each quasar
image by masking out companions and assigning the error at each radius
to be the standard deviation $\sigma$ in circular annuli centered on
the quasar.  We then re-ran the fits with the $1/\sigma^2$ weight
map and the high signal-to-noise PSF0026 (we found that the
$1/\sigma^2$ weighting scheme fails with fainter PSFs where the fits
become dominated by the noisy wings).  As
in MDK, maps of $\chi^2_\nu$ generated by the PSF0026 fits are very
uniform, so that no one area dominates the fitting.  
However, the fits generally resulted in overall $\chi^2$ values less
than the number of degrees of freedom, indicating that we have
overestimated the size of the errors.  This is because azimuthal
structure in the PSF inflates the rms error around each annulus.

We have compared the resulting galaxy parameters to those generated by
the best unweighted fits.  Excluding PG1121+422,
which is again suspect, and KUV18217+6419, for which no unweighted
deVaucouleurs fit converged even though it is clearly an $r^{1/4}$
host based on the 1D profile, the host magnitudes and axial ratios
have exactly the same spread as the uncertainties quoted above.  The
galaxy sizes, however, are significantly different.  The weighted fits
give systematically larger ellipticals and spirals by 0\farcs6 (or 1.5
times) and 0\farcs3 (or 1.9 times) respectively, with a large spread.  
One possible reason is that down-weighting the nuclear region allows
less of the point-source component, which has a small scale length, to
be attributed to the host.

\section{Discussion}\label{sec-2dresults}

\subsection{Host Galaxy Types}

One of the most interesting results of quasar host galaxy research in
the past few years has been the debunking of the textbook myth
``radio-loud quasars live in ellipticals, radio-quiet quasars live in
spirals.''  While early ground-based studies had shown this statement
to be an oversimplification (\cite{hut95}), the improvement of
instruments and the arrival of HST has made more detailed analyses
possible.  Various studies over the past few years, each
including one to several dozen nearby quasars, have claimed that
radio-quiet quasars often inhabit non-spiral hosts, and that the
elliptical fraction likely increases with nuclear luminosity.  

There have been many techniques used to assess the host type.  
McLeod \& Rieke (1995) combined their ground-based near-IR host 
magnitudes with visual inspection of WFPC2 images from several groups
to determine that there is a strong preference for high-luminosity
quasars to have smooth, early-type hosts.
Taylor et al. (1996) applied a 2D $\chi^2$ modeling technique like
that of MDK to their ground-based, near-IR images and found that
almost half of the radio-quiet quasars lie in deVaucouleurs galaxies.
R\"onnback et al. (1996) applied a $\chi^2$ test to 1D profiles to
determine that the radio-quiet objects are found in both ellipticals
and spirals; however, they performed the fits to profiles that had
already had the PSF removed, the normalization of which can alter
their results for reasons they outline.
Bahcall et al. (1997) took advantage of the superior resolution of
WFPC2 to carry out visual morphological classification of
their sample of 20 quasars, and found that more of
their radio-quiet objects appeared to be in smooth hosts than in spirals.  
Although smoothness is not proof of an elliptical, the galaxies are
also generally round, unlike the hosts of lower-luminosity quasars
imaged with WFPC2 by Hooper, Impey, \& Foltz (1997).  
Boyce et al. (1998) arrived at a similar conclusion for quasars imaged
with the WFPC2 PC, based on both morphological and cross-correlation
analyses; for the latter, they selected the model that gave the
smaller $\chi^2$.

All of the techniques have limitations, and all are based on the
possibly faulty premise that hosts are necessarily one type or the
other.  Nonetheless, we have attempted to
discern the galaxy type for the quasars in our (radio-quiet) sample by
several methods, the results of which are summarized in Table
\ref{tab-types}.
First, we have examined the images and residual maps for evidence of
spiral structure.  Although any star-forming regions will admittedly
be harder to see in the near-IR than in the visible, we see clear
evidence for arms/tidal features in only PG1322+659 and PG1048+342
(which also has a 
spiral companion), and probably PG0947+396.  
Second, we have used the profiles in Fig. \ref{fig-profs} to
determined by inspection which curve in each pair (plotted vs. $r$ or
$r^{1/4}$) better approximates a straight line.  We have shown various
normalizations for the PSF subtraction so the reader can judge the
reliability of this technique.
Third, we have used the merits from our various 2D fits, where we
record the preferred type when the difference in merit is at least 10\%.
Unfortunately, {\it even though the galaxy parameters are little
affected, the preference for one type over the other depends on the
weighting and masking used.}  Fourth, we apply the MDK $\chi^2$ test.

\begin{deluxetable}{lllllll}
\scriptsize
\tablecaption{Host Galaxy Types\tablenotemark{a}\label{tab-types}}
\tablehead{
 &  & \multicolumn{4}{c}{2D fits} \\
\cline{4-7} \\
& \colhead{Spiral/tidal} & \colhead{1D} &
\colhead{no mask} & \colhead{mask} & \multicolumn{2}{c}{$\chi^2$ weight} \\
\colhead{ID} & \colhead{arms} & \colhead{profile} & \colhead{no weight} & \colhead{no weight} & \colhead{MDK\tablenotemark{b}} &\colhead{JW\tablenotemark{c}}
}
\startdata
   PG0026+129  & d  &   d  & -  & - &  e &   -    \\
   PG0947+396  & e? &   e  & -  & e &  e &   -    \\
   PG1048+342  & e  &   d  & d  & d &  d &   d    \\
   PG1121+422\tablenotemark{d}  & d  &   d  & -  & - &  d &   -    \\
   PG1151+117  & d  &   e  & d  & - &  e &   -    \\
   PG1322+659  & e  &   -  & -  & - &  d &   -    \\
   PG1352+183  & d  &  d?  & -  & e &  e &   e    \\
   PG1354+213  & d  &   d  & d  & e &  d &   d    \\
   PG1427+480  & d  &   d  & d  & e &  d &   -    \\
   PG2233+134  & d  &   d  & -  & - &  d &   -    \\
      MARK876  & d  &  d?  & d  & e &  d &   -    \\
        UM357  & d  &  d?  & -  & - &  e &   -    \\
    Q0530-379  & d  &   d  & -  & - &  e &   -    \\
   NAB1612+26  & d  &   d  & -  & - &  e &   -    \\
  1628.6+3806  & d  &   d  & -  & - &  e &   -    \\
 KUV18217+6419  & d  &   d  & -  & - &  - &   -    \\

\enddata
\tablenotetext{a}{e=exponential, d=deVaucouleurs}
\tablenotetext{b}{$\chi^2$ criterion from McLure, Dunlop, \& Kukula (1999)}
\tablenotetext{c}{$\chi^2$ criterion from Jahnke \& Wisotzki (2000)}
\tablenotetext{d}{Detection is uncertain}
\end{deluxetable}

One might be tempted on mathematical grounds to trust exclusively the
$\chi^2$ statistic used by Taylor et al. (1996) and MDK.  Indeed, we
find by their method a result similar to theirs, that roughly half of
the radio-quiet objects are in deVaucouleurs galaxies.  Their
criterion for preferring a galaxy type was that $|\chi^2_{deV} -
\chi^2_{exp}| > C$, where $C$ is the value of $\chi^2$ that gives
99.99\% probability for the number of parameters in the fit.  However,
for reasons outlined by Jahnke \& Wisotzki (2000), this too is
problematic.  We therefore also consider the latter authors' more
conservative $\chi^2$ criterion, namely that a galaxy type is preferred if it
cannot be ruled out at 95\% confidence, while the alternate model is
rejected, say at the 99.9\% level (the results are insensitive to the
exact level chosen).  By this criterion, we make a clean distinction
in only {\it one} case: PG1352+183 is an exponential.  As discussed above, we
have likely overestimated the errors in many cases.  We therefore also
rescale the errors to achieve $\chi^2_\nu=1$ and repeat the test.
This adds to the list PG1048+342, with a deVaucouleurs host, and
PG1354+213, with an exponential host.  Interestingly, even by the
conservative criteria, we find an exponential host with no spiral
arms, and a deVaucouleurs host that has arms or tidal features.

Table \ref{tab-types} shows that the methods used by various authors
to discriminate the host types give discrepant results, and hints that all
such result should be treated with caution.  At least in our sample,
exponentials and deVaucouleurs models are nearly equally good (or bad)
at fitting the 2D images.  As the reader can see from Figures 
\ref{fig-qimages} and \ref{fig-profs},
neither model is generally perfect, and
the results can be sensitive to the fraction of nuclear light
attributed to the galaxy. 
We note that the methods that down-weight
the nuclear region, namely the masked and $\chi^2$ methods, generate a
higher fraction of disk systems.  One possible explanation is that
these are disk galaxies with significant bulge components.  When the
central regions are neglected, the fit is dominated by the outer,
disky components.  Unfortunately, with the bulges expected to be
comparable to the PSF size, we do not feel we can justify adding a
bulge component to the disk fits.  
However, the facts that arms/tidal features are visible in
only two or three of the galaxies, and that all techniques indicate a
high fraction of non-disk systems, together {\it support the notion that
many of the radio-quiet quasars live in non-spiral galaxies}.  

For completeness, we also examined the distributions of host-galaxy
axial ratios $b/a$.  For both exponential and deVaucouleurs models,
there is a lack of hosts with low $b/a$.  This is suggestive of an
early-type population, but could also be the result of a selection
effect.  At least for 
low luminosities, there is a strong bias against finding active nuclei
in edge-on spirals, where obscuration in the plane of the host obscures
the nucleus (\cite{sim97}).

\subsection{Interactions}

Interactions have long been suspected of triggering and fueling of
quasars, and HST images of nearby quasars have revealed some
spectacular examples of hosts with tidal distortions
(e.g. \cite{hut94}; \cite{bah97}; \cite{boy99}; \cite{mcl99}).  
In our sample, the fraction of hosts that appear to be involved in strong
interactions (approximately 4/16) is similar to what has been found in
previous studies.  However, approximately half of the objects appear
to have hosts that are smooth and symmetrical at HST resolution.  This
is a lower rate than in other studies, though the small numbers make
the difference insignificant.  For a more detailed discussion of the
interaction levels of the quasars from our high-luminosity sample
compared to those of low-luminosity quasars, see McLeod \&
Rieke 1994b.

\subsection{Eddington Fractions and the Luminosity/Host-Mass Limit}

McLeod \& Rieke (1995a), defined a luminosity/host-mass limit for
quasar hosts based on a compilation of ground-based near-infrared data. 
This is not a correlation, as recently misinterpreted by some groups,
but rather a limit on the maximum nuclear luminosity possible in a
host of a given near-IR luminosity (and presumably mass).  From the
ground-based data, we found a tight limit well-described by
$M_B(nuc)\approx M_H(host)$.  

MRS interpret the limit in terms of physical parameters and provide a
formula for calculating the fraction of the Eddington luminosity
$L/L_{edd}$ at which the quasar radiates.  The calculation uses the
nuclear B magnitude and the host-galaxy H magnitude along with
assumptions about the galaxy mass-to-light ratio, the fraction of
the galaxy's mass that resides in the central black hole, and the
energy distribution of the nucleus.   Under these assumptions, the
original $M_B(nuc)\approx M_H(host)$ line translates to a constant
Eddington fraction of $\approx 15\%$.  

\begin{figure}
\epsscale{0.6}
\plotone{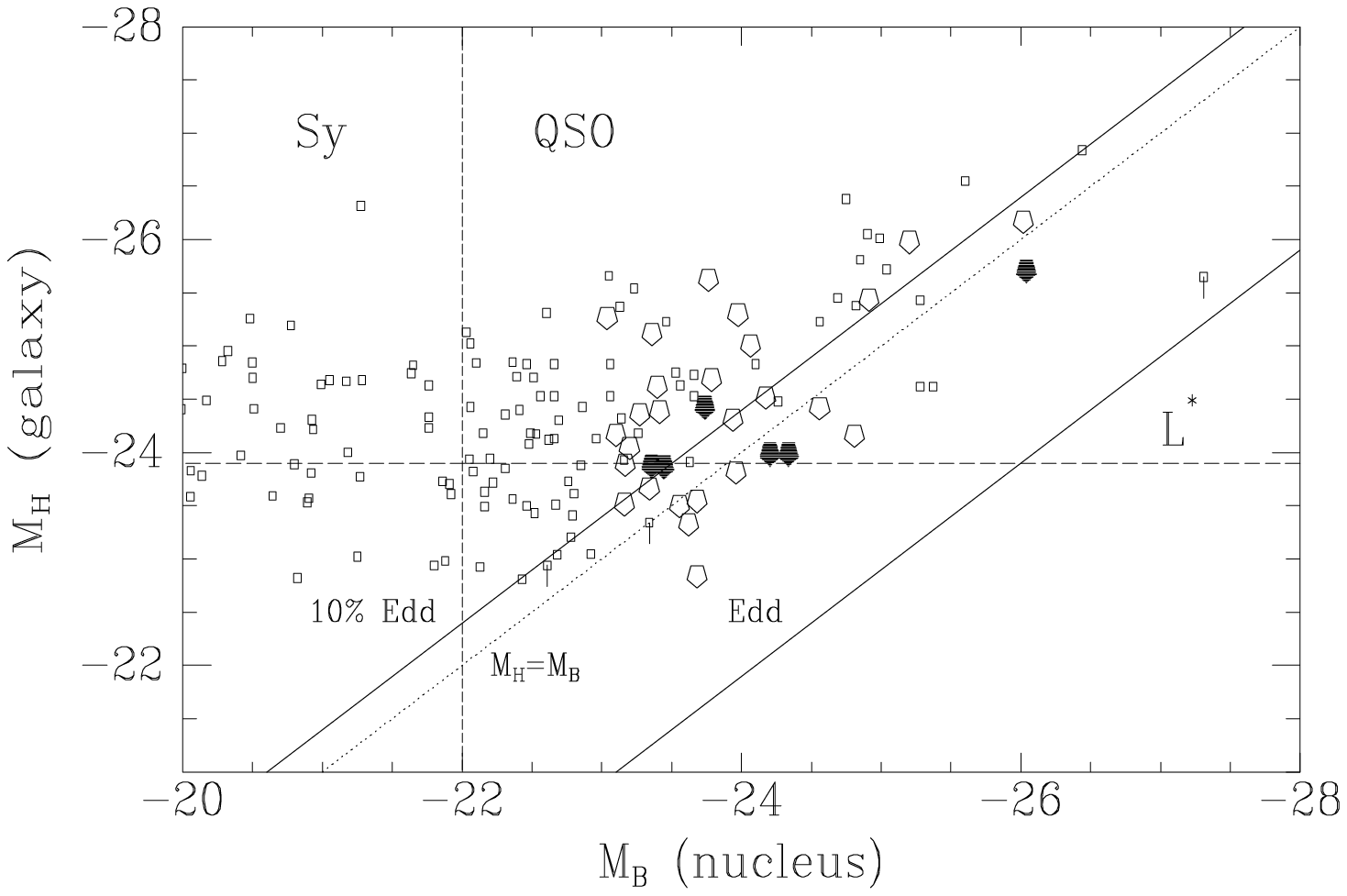}
\figcaption[mags.eps]{Galaxy versus nuclear absolute magnitudes for QSOs.
Low-redshift QSOs and Seyferts shown as boxes are taken from
MRS and references therein.  QSOs shown as open pentagons
constitute the high-luminosity sample from \cite{mr94b}, with host
magnitudes derived from either NICMOS images (this paper)
or WFPC2 images (\cite{bah97}; \cite{mcl99}).  Filled pentagons are the
other 6 QSOs from our NICMOS imaging.  Also shown are the
QSO/Sy boundary (dashed vertical line), position of an $L^*$ galaxy
(dashed horizontal line), and loci of Eddington and 10\% Eddington
luminosities.  The original ``luminosity/host-mass limit'' is shown by
dotted line for reference.\label{fig-mags}}
\end{figure}

The data for the current project were taken to probe further the
luminosity/host-mass limit, and to try intentionally to find galaxies
that violate the limit.  In Figure \ref{fig-mags} we show the host and
nuclear magnitudes based on the 1D analysis in this paper.  Owing to
an updated F160W zero point, more accurate saturation limits, and an
improved 1D PSF-subtraction technique, the host magnitudes in the
current paper do differ, sometimes substantially, from those in MRS. 
The zero point accounts for 0.31~mag of the difference.  
The saturation limit has contributed to a major revision downward of
the host magnitude for KUV1821+643, which Percival et al. (2000) noted
had an unrealistically luminous host in MRS.  
The improved 1D technique has corrected a tendency for the host flux
to be overestimated in quasars with the lowest ratio of
host-to-nuclear light (especially the 6 objects we picked specifically
because their hosts had not been detected from the ground, represented
by the black pentagons in the Figure). 

The magnitudes derived for our sample from HST data make the limit 
less of a straight line in mag-mag space, but they reinforce the basic
result that the most luminous quasars are not found in low-luminosity
galaxies.  We list in Table \ref{tab-1d} the  
Eddington fractions derived from the current 1D analysis, and note
that the results from the 2D deVaucouleurs analyses are similar.  The
basic MRS conclusion holds, namely that the Eddington fractions are
generally $<20\%$.  In fact, the only quasar exceeding this limit in
the present study is PG1121+422, with $L/L_{edd}=0.31$, but our
detection of this galaxy is suspect.  Its host could be 3 times
fainter without crossing the Eddington line.

Recently, Percival et al. (2000) have probed the limit by observing at K
the hosts of 14 high-luminosity, radio-quiet quasars.  They find that
their galaxies lie ``to the right of the McLeod \& Rieke points'' on a
plot of nuclear absolute magnitude versus host absolute magnitude,
i.e. at higher nuclear luminosity for a given host luminosity than the
hosts in our ground-based studies.  As shown in Figure \ref{fig-mags},
some of our quasars do also.
Percival et al. (2000) also state that our limit is difficult to
transfer quantitatively to their Figure 5 (the PG Survey B
magnitudes we used were shown to be systematically too bright by
Goldschmidt et al. (1992), but only by $\sim0.3~\rm mag$; the color
transformations from R to B and H to K are sensitive to the assumed
spectra; etc.).  However, we can make an approximate comparison by
noting that on the MRS plot, the Eddington line crosses the $L^*$ host galaxy
line two nuclear magnitudes brighter than the envelope described by
our ground-based limit.  This is precisely the region occupied by the
Percival et al. (2000) quasars.  We conclude that the Percival et
al. (2000) quasars also likely are radiating below the Eddington
limit, though perhaps closer to it.  This is not surprising
given that they targeted the most luminous nuclei.  Further evidence
for consistency with the limit can be seen from their Figure 5, where
their quasars overlap with those that McLure et al. (1999) have shown
to be sub-Eddington radiators. 

A re-statement of the luminosity/host-mass limit based on the physical
model is to say that quasars today typically radiate below $\approx
20\%$ of the Eddington rate.   Despite choosing quasars that have high
nuclear luminosity, and ones for which ground-based attempts to image
the hosts had failed, we have not yet found any object that violates
the limit corresponding to $L=L_{edd}$.  However, we expect such
object will be found at higher redshifts according to the predictions
of Kauffmann \& Haehnelt (2000).  Their hierarchical galaxy formation
models imply that the massive hosts of today's highest-luminosity
quasars formed relatively recently, but that at earlier epochs,
luminous quasars will be found in progressively lower-luminosity hosts.

\subsection{Notes on Individual Objects}

The following notes are based on inspection on an image display of Fig.
\ref{fig-qimages} as well as our full (20\arcsec) reduced
images.  The magnitudes of other objects in the fields are simple
aperture magnitudes.

\begin{itemize}

\item[\bf PG0026+129]{Smooth host with no distortions visible in
PSF-subtracted image.  A galaxy with $H=19.2$ is seen at 4\farcs7
separation, and a faint compact object is at 3\farcs1.} 

\item[\bf PG0947+396]{Possible broad spiral structure visible in
PSF-subtracted image.  This may be tidal in origin, due to an
interaction with the large edge-on spiral galaxy with $H=16.65$ seen
at 9\farcs9.  If at the redshift of the quasar, this spiral is at a
projected separation of 28~kpc. 
A second galaxy with $H=18.7$ is at 5\farcs0.}

\item[\bf PG1048+342]{Spiral structure visible in PSF-subracted
image, but the 1D profile indicates a deVaucouleurs profile.  As with
PG0947, the structure could be tidal in origin, due to interaction
with the large spiral ($H=17.3$, and similar in size to the quasar host)
at a projected separation of 3\farcs4 (8.2~kpc if at the quasar's redshift).
This galaxy appears to be part of a group.  Besides the large spiral,
there are at least 5 fainter spirals and distorted galaxies within 18\arcsec.
}

\item[\bf PG1121+422]{No host is visible on the images, but the 1D
profile suggests that a compact one is present.}

\item[\bf PG1151+117]{Smooth host is visible on the PSF-subtracted
image.  DeVaucouleurs fit residual shows an elongated structure
right-left (north-south).
An $H=18.8$ galaxy is at 3\farcs6, and a faint
point source is roughly twice that far away.}

\item[\bf PG1322+659]{Beautiful 2 armed spiral host with bar-like
structure right-left (north-south).
Several galaxies, including one that is likely similar in size to the
host, are visible on the edge of our full frame at $\sim10\arcsec$.
If at the redshift of the quasar, the large galaxy is at a projected
separation of 24~kpc and could be responsible for the host's broad arms.} 

\item[\bf PG1352+183]{Smooth host.
A faint ($H=21$), elongated galaxy is seen at 5\farcs2.}

\item[\bf PG1354+213]{Smooth host.}

\item[\bf PG1427+480]{Smooth host.  An $H=19.6$ galaxy is at
3\farcs1.}

\item[\bf PG2233+134]{Smooth host.  There are two small galaxies in
our full frame: $H=20.6$ at 3\farcs6, $H=19.6$ at 9\farcs7.}

\item[\bf MARK876]{[Also called PG1613+658] Severely distorted host.
The host galaxy filled so much of the frame that removal of the NICMOS
quadrant effect was problematic.  As a result, the frame is not as flat as
for the other objects.  However, the nearly triangular outer contours
with a straight left edge is not an artifact, but is also visible on
ground-based images.  What the NICMOS image reveals is that a companion,
with $H=16$ and at a projected separation of 2\farcs2 (4.3~kpc), is
appears to be embedded within the host.  The PSF-subtracted image
shows large-scale, asymmetric, tidal features that probably resulted
from this interaction.  
} 

\item[\bf UM357]{Host is very compact, and too small for structure to
be visible.}

\item[\bf Q0530-379]{Smooth host.  There are several other galaxies
visible in the full frame:  $H=19.4$ at 4\farcs2, $H=19.8$ at
6\farcs5, $H=18.3$ at 8\farcs7, and $H=20.6$ at 9\farcs7.}

\item[\bf NAB1612+26]{Host is compact but clearly elongated
left-right (northwest-southeast).  Two diffuse, faint galaxies are
visible in our full frame: $H=18.8$ at 6\farcs0, and $H=21.6$ at 9\farcs2.
}

\item[\bf 1628.6+3806]{PSF-subtracted image shows asymmetric extension
to lower left (northeast).
There are two small galaxies in the frame: $H=19.3$ at
3\farcs5, and $H=20.5$ at 5\farcs6.  There is also a star at 5\farcs6.}

\item[\bf KUV18217+6419]{1D profile shows that this is clearly a
deVaucouleurs host, yet deVaucouleurs fits converged only when masked.
This is perhaps due to interference from a linear feature extending 
up (southwest) from the host.
However, it is also possible that the extremely high nuclear
brightness has compromised the fitting; it has by far the brightest
nucleus in our sample, and the host-to-nucleus flux ratio is low.
Still, the galaxy is intrinsically the most luminous in our sample.
There are two galaxies at 9\farcs5 ($H=17.3,~17.6$) seen on opposite sides
of the quasar in our full frame. 
}

\end{itemize}

\section{Conclusions}

We have imaged 16 low-redshift, high-luminosity quasars with NICMOS.
Using stellar images as PSFs has worked extremely well, and NICMOS has
done a superb job of showing the host galaxies with high contrast
against the bright nucleus.  We find the following results.   

\begin{itemize}

\item [(i)] For redshifts $z\lesssim0.3$, the host-galaxy magnitudes
derived from ground-based data using our 1D analysis technique are very 
reliable (within $\sim 0.30~\rm mag$) unless there are companions that
cannot be resolved.  Therefore, general conclusions about host luminosities
from our previous papers should be robust. 

\item [(ii)] Distinguishing between deVaucouleurs and exponential
hosts based on various fitting techniques is questionable, especially
given that neither law is likely perfect.  For 2D fits, we recommend
a very conservative $\chi^2$ criterion for discrimination.  While some
previously used methods appear to distinguish between types in almost
every case, they are often at odds with the more conservative
criterion, which can distinguish between types in only $\sim3$ cases
out of 16.  The 1D radial profiles are also very useful.  Fortunately,
HST allows us to look directly for spiral arms and other morphological
features in the hosts.  We concur with previous studies that have
found radio-quiet quasars often live in deVaucouleurs hosts.  

\item [(iii)] Approximately 4 of the 16 hosts are undergoing
strong interactions with companions.  However, nearly half of the
hosts are smooth and symmetric, a reminder that current-epoch
interactions are not a necessary condition of quasar activity.

\item [(iv)] Assuming that galaxies contain central black holes with
0.6\% of the galaxy's mass (see MRS for details), 15 of the 16 quasars
in our sample radiate at $\sim2-20\%$ of the Eddington rate.  Our host
detection for PG1121+422 is uncertain, so its rate of 31\% Eddington
may be a lower limit.  Despite intentionally choosing high-luminosity
quasars whose hosts were hard to detect from the ground, we have
failed to find any object that violates a luminosity/host-mass
limit corresponding to $L=L_{Edd}.$  

\end{itemize}

\acknowledgments Thanks to George Rieke and Lisa Storrie-Lombardi for 
their work on the HST proposal and MRS, and to Lisa also for
especially valuable comments on this paper.   We thank the NICMOS team
for putting such a superb instrument in orbit, Erin Condy for help
with the data reduction, Ray Weymann for helpful ideas
during the course of this project, and Marcia Rieke for the
calibration files used with $nicred\_1.8$.  We thank the anonymous
referee for comments that helped us to improve the
presentation significantly. 
Support for this work was
provided by NASA through grant number GO-07421.01-96A from the Space
Telescope Science Institute, which is operated by the Association of
Universities for Research in Astronomy, Inc., under NASA contract
NAS5-26555.  We also gratefully acknowledge support from the Keck
Northeast Astronomy Consortium.

\end{document}